# A light-weight full-band speech enhancement model

*Qinwen Hu[1,a,b] , Zhongshu Hou[1,a,c] , Xiaohuai Le[1,d] , Jing Lu[1,e]*

[1] Key Laboratory of Modern Acoustics, Nanjing University,
Nanjing, 210093, China

**Abstract:** Deep neural network based full-band speech enhancement systems face challenges of high demand of computational resources and imbalanced frequency distribution. In this paper, a light-weight full-band model is proposed with two dedicated strategies, i.e., a learnable spectral compression mapping for more effective high-band spectral information compression, and the utilization of the multi-head attention mechanism for more effective modeling of the global spectral pattern. Experiments validate the efficacy of the proposed strategies and show that the proposed model achieves competitive performance with only 0.89M parameters.

## 1. INTRODUCTION

In the last decade, data-driven speech enhancement (SE) approaches based on Deep Neural Networks (DNN) have achieved remarkable progress [1] with more noise suppression and higher speech quality than the conventional signal processing methods. Most of the SE systems focus on wide-band (16 kHz) speech, while full-band (48 kHz) SE methods are still to be explored for their implementation in scenes that require high-fidelity audios.

Directly expanding the bandwidth and applying the wide-band processing network to full-band SE would result in significant increases of the memory requirement and computational burden and thus impeding its implementation in portable devices with limited computational resources. Furthermore, processing all the frequency points in a uniform manner is inefficient, because most of the human speech energy and tonality information concentrate in the low frequency range [2]. Spectral envelopes can be used to compress the features [3][4], but the low resolution restrains the network's performance. Sub-nets optimized for different bands [5] can focus on the low frequency processing in a flexible manner and significantly improve SE performance. However, the whole spectrum still needs to be processed, making it hard to shrink the model size.

In this paper, we propose a light-weight full-band SE model based on our recently proposed dual-path convolutional recurrent network (DPCRN) [6] for wide-band SE. Ranked at the 3rd place in Deep Noise Suppression-3 (DNS-3) challenge [7], it has much fewer parameters and lower computational burden than the other top models (0.8M parameters and 3.7G MACs, compared with 6.4M parameters and 6.0G MACs of Rank-1 model and 5.2M parameters and 6.0G MACs of Rank-1 model and 5.2M parameters and 52.5G MACs of Rank-2 model), which makes it a very competitive light-weight SE method. To better exploit the inherent connection among the spectrum, an attention mechanism is introduced into our proposed model, i.e., the recurrent neural networks (RNN) for the intra-chunk processing are replaced by multi-head attention (MHA) networks [8] while the RNNs for the inter-chunk processing are retained, so the proposed model is named as dual-path attention-recurrent network (DPARN). To more efficiently process the high frequency components, where the speech is sparsely distributed, a dedicated learnable spectral compression mapping (SCM) and its inversion are added to the preprocessing and postprocessing respectively, so that the size of the whole model can be effectively compressed. Our model achieves competitive performance compared to other state-of-the-art (SOTA) full-band methods with only 0.89M parameters.

## 2. MODEL DESCRIPTION

### 2.1. Problem formulation

The time-frequency (T-F) domain noisy speech can be described as
$$X(t,f) = S(t,f) + N(t,f), \qquad (1)$$
where $t$ denotes the time frame and $f$ denotes the frequency, which are omitted when we refer to the whole spectrogram.

The target of our proposed model is to map the input noisy speech $X$ to the estimated clean speech $\tilde{S}$ and minimize the loss function $L$ calculated on $\tilde{S}$ and the ground-truth clean source $S$,
$$\tilde{S} = \mathcal{F}(X), \qquad (2)$$
$$L = Loss(\tilde{S}, S), \qquad (3)$$
where $\mathcal{F}$ refers to the network function that acts on the complex spectrum. We use a mapping-based method to directly estimate the real and imaginary parts of the target spectrum, which can restore high-band spectrum of the original speech even when the input signal is low-pass filtered.

### 2.2. Learnable spectral compression mapping

Directly expanding the wide-band network to full-band processing with the same spectral resolution would allocate two thirds of the computational resources to the less informative high band (8 kHz-24 kHz), resulting in significant increases of both the computation-computational resources to the less informative high band (8 kHz-24 kHz), resulting in significant increases of both the computation-

[a] Contribute equally to this work.
[b] qinwen.hu@smail.nju.edu.cn
[c] zhongshu.hou@smail.nju.edu.cn
[d] xiaohuaile@smail.nju.edu.cn
[e] lujing@nju.edu.cn

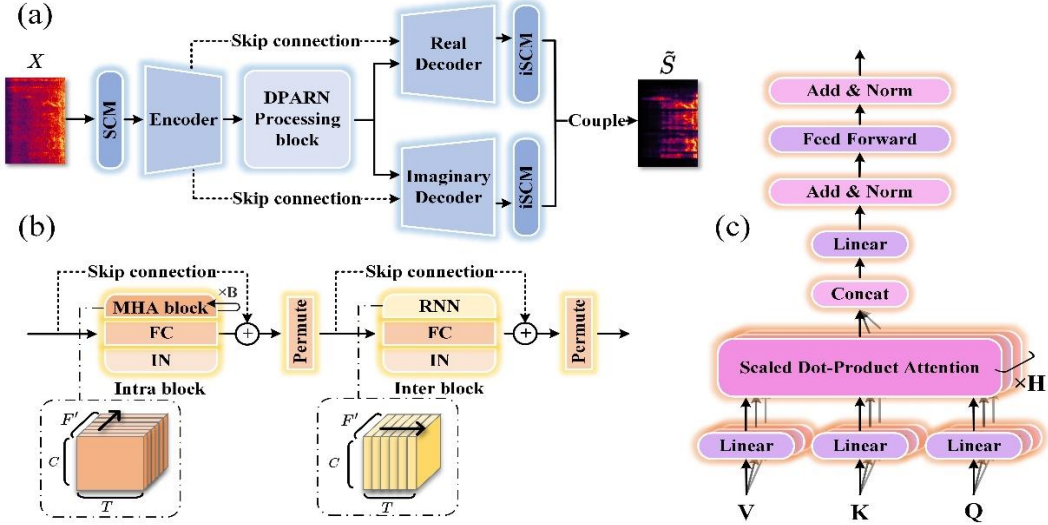

Figure 1: (a) The overall structure of DPARN. (b) The structure of the DPARN processing block, where the chunks in the dashed boxes show how the sequential modeling is implemented in the intra-block and the inter-block, and the arrows indicate the directions of the modeling. (c) The multi-head attention block.

al burden and the learning difficulty. Therefore, it is necessary to introduce a spectral information mapping strategy to effectively compress the high band. The SCM layer is designed following similar rules of the Mel-scale filter banks, which convert frequencies into Mel-scale by a logarithmic function [9][10]. To further retain the information in the critical low band, we warp the spectrum by keeping frequencies below 5 kHz intact while only transforming frequencies above 5 kHz logarithmically, with the mapping curve defined as

$$f_c(m) = \begin{cases} f(m), & 0 \leq f(m) \leq 5\text{ kHz} \\ 2500\left[\ln\left(\frac{f(m)-2500}{2500}\right)+2\right], & 5\text{ kHz} < f(m) \leq 24\text{ kHz} \end{cases} \quad (4)$$

where $m$ denotes the index of frequency bin. The $F$-dimensional spectrum can be compressed into an $F_c$-dimensional spectral representation with $f_c(k)$ ($k = 1,2,...,F_c$) uniformly spaced in the transformed domain, and the transformation can be described as

$$\boldsymbol{x}_{comp} = W_{SCM}\boldsymbol{x}, \boldsymbol{x} \in \mathbb{R}^F, \boldsymbol{x}_{comp} \in \mathbb{R}^{F_c}, \quad (5)$$

$$W_{SCM} = \begin{bmatrix} \boldsymbol{I}_{K \times K} & \boldsymbol{0}_{K \times (F-K)} \\ & \boldsymbol{G} \end{bmatrix} \in \mathbb{R}^{F_c \times F}, \quad (6)$$

$$\boldsymbol{G} = \left[\boldsymbol{g}_{K+1}, \boldsymbol{g}_{K+2}, ..., \boldsymbol{g}_k, ..., \boldsymbol{g}_{F_c}\right]^T \in \mathbb{R}^{(F_c-K) \times F}, \quad (7)$$

$$\boldsymbol{g}_k(m) = \begin{cases} 0, & f(m) < f(f_c(k-1)) \\ \frac{f(m)-f(f_c(k-1))}{f(f_c(k))-f(f_c(k-1))}, & f(f_c(k-1)) < f(m) \leq f(f_c(k)) \\ \frac{f(f_c(k+1))-f(m)}{f(f_c(k+1))-f(f_c(k))}, & f(f_c(k)) < f(m) \leq f(f_c(k+1)) \\ 0, & f(m) > f(f_c(k+1)) \end{cases}$$

$$(m = 1,2,...F; k = K+1, K+2, ..., F_c), \quad (8)$$

where $\boldsymbol{x}$ is the original spectrum, $\boldsymbol{x}_{comp}$ is the transformed representation, $K$ is the index corresponding to the 5-kHz threshold, $\boldsymbol{g}_k$ is the $k$th triangular filter, $f(m)$ is the physical frequency of the $m$th bin in the original spectrum, and $f(f_c(k))$ is the inverse mapping of the logarithmic function in (4), i.e.,

$$f(f_c(k)) = 2500\left(e^{\frac{f_c(k)}{2500}-2} + 1\right).$$

Note that the mapping of (4) roots from the human auditory system and cannot effectively match the sparse distribution of speech spectrum in high frequency range. Therefore, the direct implementation of this compression pattern would lead to considerable residual noise.

To more effectively exploit the sparse distribution of speech in the high-frequency range, we propose to use a partially learnable compression matrix initialized by $W_{SCM}$. The matrix is learned as the weights $\widetilde{W}_{SCM}$ of a fully-connected (FC) layer with no biases. The low-band mapping $\widetilde{W}_{SCM}^{low} = [\boldsymbol{I}\ \boldsymbol{0}] \in \mathbb{R}^{K \times F}$ is fixed, and the high-band part $\widetilde{W}_{SCM}^{high} \in \mathbb{R}^{(F_c-K) \times F}$ is learned by the network, which is initialized by $\boldsymbol{G}$ in (7). Correspondingly, the inverse spectral compression mapping (iSCM) is also realized through a learnable matrix $\widetilde{W}_{iSCM} \in \mathbb{R}^{F \times F_c}$, but with random initialization.

### 2.3. Dual-Path Attention-Recurrent Network

Similar to DPCRN [6], the proposed model consists of an encoder, a dual-path processing block and two decoders, as depicted in Figure 1(a). The encoder follows the SCM layer and contains multiple 2D convolutional layers, while every decoder is followed by an iSCM layer and contains multiple transposed-2D convolutional layers to reconstruct the real or imaginary part of the spectrum. Skip connections are utilized between the corresponding layers in the encoder and decoders.

In the original DPCRN, intra-chunk RNNs are applied to model the correlations among frequencies in a single frame, and inter-chunk RNNs are applied to model the time dependence of a certain frequency. Considering the significantly wider frequency span of full-band speech, we replace the intra-chunk RNNs with MHA [8], because it can more effectively model the long-term global spectral patterns. On the other hand, global information along the time axis is not essential for SE, so the inter-chunk RNNs are retained.

Figure 1(b) shows the detailed structure of the DPARN processing block. $C$ dimensional local features are extracted by the encoder for every bin in the compressed spectrogram. Trigonometric positional encodings [8] (PE) are added to the input of intra-MHA to encode the positional information. The queries ($Q \in \mathbb{R}^{F' \times C}$), keys ($K \in \mathbb{R}^{F' \times C}$), and values ($V \in \mathbb{R}^{F' \times C}$) are all replicas of the attention layer input, where $F'$ denotes the frequency dimension squeezed by SCM and the convolutional encoder. The attention mechanism is conducted with $H$ parallel heads. In each head, $Q$, $K$ and $V$ are linearly projected to $d_q$, $d_k (d_k = d_q)$, and $d_v$ dimensions respectively as:

$$Q_h = QW_h^Q, K_h = KW_h^K, V_h = VW_h^V,$$

$$(W_h^Q \in \mathbb{R}^{C \times d_q}, W_h^K \in \mathbb{R}^{C \times d_k}, W_h^V \in \mathbb{R}^{C \times d_v}) \quad (9)$$

where $h$ is the head index. The scaled dot-product attention is applied on $Q_h$, $K_h$ and $V_h$ afterwards as

$$Attention(Q_h, K_h, V_h) = softmax\left(\frac{Q_h K_h^T}{\sqrt{d_k}}\right) V_h, \quad (10)$$

which calculates the similarities between the projected representations of each frequency pair and assigns significance to the projected value $V_h$ accordingly. The attention output of different heads is then concatenated and linearly projected back to a series of $C$ dimensional vectors as

$$MHA(Q, K, V) = Concat(head_1, head_2, \ldots, head_H)W^O,$$

$$head_h = Attention(Q_h, K_h, V_h), W^O \in \mathbb{R}^{(H \times d_v) \times C}, \quad (11)$$

The graphical illustration of MHA is illustrated in Figure 1 (c).

Following the MHA layer, a feed-forward network further processes the information at each frequency point, which includes two FC layers with a ReLU function as

$$FFN(x) = max(0, xW_1 + b_1)W_2 + b_2,$$

$$(W_1 \in \mathbb{R}^{C \times (4C)}, W_2 \in \mathbb{R}^{(4C) \times C}, b_1 \in \mathbb{R}^{4C}, b_2 \in \mathbb{R}^C). \quad (12)$$

One MHA layer and one feed-forward network are regarded as a single MHA block. An FC layer and an instance normalization (IN) follow $B$ repetitive MHA blocks. The chunks are then permuted and fed into the inter-RNN block. In the inter-RNN block, a long short-term memory (LSTM) layer is used, which is also followed by an FC layer and an IN. Skip connections are applied within each MHA block [8] and between the inter- and intra- blocks [6].

## 2.4. Training (or Learning) target

The learning target is the real and imaginary parts of the clean speech spectrogram $S = S_{real} + iS_{imag}$, where $S_{real}$ denotes the real part and $S_{imag}$ denotes the imaginary part. A power compress loss function [11] is used to better process the information in low power T-F points. The complex spectrogram can be written as $S = |S|e^{i\theta_S}$ in polar coordinates, and then the power-compressed spectrum can be described as $S^C = |S|^\gamma e^{i\theta_S}$, where $\gamma$ refers to the compression parameter and the superscript $C$ denotes the power compressed pattern.

The real and imaginary parts are thus given by

$$S_{real}^C = |S|^\gamma \cos\theta_S, \ S_{imag}^C = |S|^\gamma \sin\theta_S. \quad (13)$$

$\tilde{S}_{real}^C$ and $\tilde{S}_{imag}^C$ of the estimated clean speech follow the same definition.

The loss functions for restoring complex spectrogram and magnitude spectrogram, namely $L_{RI}(\tilde{S}, S)$ and $L_{Mag}(\tilde{S}, S)$, can then be described as

$$L_{RI}(\tilde{S}, S) = \left\|S_{real}^C - \tilde{S}_{real}^C\right\|_F^2 + \left\|S_{imag}^C - \tilde{S}_{imag}^C\right\|_F^2, \quad (14)$$

$$L_{Mag}(\tilde{S}, S) = \left\||S|^\gamma - |\tilde{S}|^\gamma\right\|_F^2, \quad (15)$$

where $||\cdot||_F$ refers to the Frobenius norm of the matrix. The loss function applied in training DPARN is

$$L = L_{RI}(\tilde{S}, S) + L_{Mag}(\tilde{S}, S). \quad (16)$$

The source code of our model is available at https://github.com/Qinwen-Hu/dparn.

## 3. EXPERIMENTS

### 3.1. Experiment I (ablation study)

#### 3.1.1. Datasets

We first conduct an ablation study to evaluate the efficacy of the SCM layer and the performance of DPARN compared with other dual-path models. We train the models on a small dataset, which consists of clean speeches from English dataset VCTK [12] and French dataset SIWIS [13], and noise data from DEMAND [14] and QUT-NOISE [15]. All these audios are sampled at 48 kHz. The size of clean speech data is around 45 hours. The audios are randomly split into 10-second long segments, and 16000 clips of clean speeches are generated in total. 14500 clips are used for training and the rest are for validation. The noisy clips are generated by first convolving 10% of clean data with room impulse responses selected from openSLR26 and openSLR28 [16], and then randomly mixing the clips with noise clips at signal-to-noise ratios (SNR) between -5 dB and 15 dB with an interval of 1 dB.

For testing, we use clean speech from DAPS [17] and noise from Saki [18]. The way we simulate our test data is the same as above. The SNR levels are {-5dB, 0dB, 5dB, 10dB, 15dB}.

#### 3.1.2. Parameter setup and training strategy

The window length for short-time Fourier transform (STFT) is 25 ms and the hop size is 12.5 ms. The fast Fourier transform length is thus 1200 points, resulting in a dimension of 601 for frequency features fed into the network. Hanning window is used when performing STFT.

In the SCM layer, $F_c = 256$, $K = 125$. The encoder contains five 2-D convolutional layers. The output channel dimensions are {16, 32, 48, 64, 80}, the kernel sizes are {(5,2),(3,2),(3,2),(3,2),(2,1)}, and the strides are {(2,1),(1,1),(1,1),(1,1),(1,1)}, where the first number refers to configuration in the frequency axis and the second number refers to that in the time axis. The transposed convolutional layers in decoders are sequenced reversely. Every (transposed) convolutional layer is followed by a batch normalization and a PReLU function. Skip connections are realized by concatenating the inputs in the channel dimension. The DPARN processing block consists of only one intra-block and one inter-block. In the intra-block, $B = 2$, $H = 8$, $d_k = d_q = d_v = C/H$ with $C = 80$; In the inter-block, the hidden size is 127.

A warm-up strategy [8] is employed to train the DPARN. The learning rate $\alpha$ varies with training steps $\psi$ as $\alpha = \frac{1}{\sqrt{C}} \times min\left(\frac{1}{\sqrt{\psi}}, \frac{\psi}{\sqrt{\Psi^3}}\right)$, and the warm-up step $\Psi$ is 40000. Adam optimizer is used with parameters $\beta_1 = 0.9$, $\beta_2 = 0.98$, $\epsilon = 10^{-9}$. The compression parameter $\gamma$ is $\frac{2}{3}$.

### 3.1.3. Baselines and evaluation metrics

We compare our proposed model with the DPCRN [6] baseline. The DPCRN model is trained with the entire 601-dimension frequency features. We also compare our model with 4 variants: DPCRN with SCM layer (DPCRN-SCM), the model using intra-RNN and inter-MHA (DPRAN-SCM), the model using intra-MHA and inter-MHA (DPAAN-SCM). $B$ is also 2 in the inter-MHA block. The evaluation metrics include perceptual evaluation of speech quality (PESQ), shorter-time objective intelligibility (STOI), and scale-invariant signal-to-distortion ratios (SI-SDR).

### 3.1.4. Results and analysis

Results for experiment I are shown in TABLE I. Without any spectral compression, DPCRN performs worse than the version with the SCM layer, supporting our argument that keeping full frequency information is not only superfluous in computation, but also detrimental to performance. The proposed DPARN-SCM achieves the best scores. Performance degradation is observed in DPRAN-SCM and DPAAN-SCM, validating that applying the multi-head attention mechanism only along the frequency axis is indeed the best choice.

TABLE I. Results for experiment I.

| Models | PESQ | STOI | SI-SDR |
|---|---|---|---|
| Noisy | 1.45 | 0.90 | 5.00 |
| DPCRN | 2.03 | 0.88 | 8.93 |
| DPCRN-SCM | 2.48 | 0.92 | 11.57 |
| DPARN-SCM | **2.65** | **0.93** | **12.56** |
| DPRAN-SCM | 2.31 | 0.92 | 10.68 |
| DPAAN-SCM | 2.10 | 0.91 | 10.29 |

## 3.2. Experiment II

### 3.2.1. Datasets

To compare DPARN with some of the latest SOTA models, we train and test our model on the open VCTK-DEMAND dataset [19]. It contains generated noisy speeches and corresponding clean speeches. The clean speeches are selected from VCTK [12], including 28 speakers for training and 2 unseen speakers for testing. The noise data includes two artificially generated (speech-shaped noise and babble) noise types and eight real noise types from DEMAND[14] for training, and five other types for testing. The SNR levels are {0dB, 5dB, 10dB, 15dB} for training and {2.5dB, 7.5dB, 12.5dB and 17.5dB} for testing. The training data is about 10 hours in total.

### 3.2.2. Parameter setup and training strategy

The STFT configuration, the network parameter setup and training strategy are the same with those in experiment I, except that the warm-up step is 5000.

### 3.2.3. Baselines and evaluation metrics

The baselines include: RNNoise [20], PerceptNet [3], DeepFilterNet [4] and S-DCCRN [5]. It should be noted that, among these models, only S-DCCRN [5] is trained on the VCTK-DEMAND training set, and it is a super-wide-band (32 kHz) model.

### 3.2.4. Results and analysis

The results for the experiment II are shown in TABLE II. Note that although RNNoise only has 0.06M parameters, its performance is significantly worse than other models. Our proposed model achieves the best scores for all three metrics with a model size of only 0.89M parameters.

TABLE II. Comparison of performance on VCTK+DEMAND dataset.

| Models | Para. (M) | PESQ | STOI | SI-SDR |
|---|---|---|---|---|
| Noisy | - | 1.97 | 92.1 | 8.41 |
| RNNoise [20] (2018) | 0.06 | 2.29 | - | - |
| PerceptNet [3] (2020) | 8.0 | 2.73 | - | - |
| DeepFilterNet [4] (2022) | 1.80 | 2.81 | - | 16.63 |
| S-DCCRN [5] (2022) | 2.34 | 2.84 | 94.0 | - |
| DPARN (2022) | 0.89 | **2.92** | **94.2** | **18.28** |

## 4. CONCLUSION

A light-weight full-band speech enhancement model called DPARN is proposed in this paper. A learnable spectral compression mapping is utilized to more effectively compress the less-informative high-band spectrum, and the multi-head attention mechanism replaces RNN for intra-chunk processing to model the global structure of full-band spectrum. With only 0.89M parameters, the proposed DPARN achieves SOTA performance compared with many recently proposed full-band SE models.

## 5. REFERENCES

[1] Wang, D. and Chen, J. (2018). "Supervised speech separation based on deep learning: An overview," IEEE Trans. Audio. Speech. Lang. Processing, 26(10), 1702-1726.

[2] Monson, B. B., Lotto, A. J., and Story, B. H. (2012). "Analysis of high-frequency energy in long-term average spectra of singing, speech, and voiceless fricatives," J. Acoust. Soc. Am. 132(3), 1754-1764.

[3] Valin, J. M., Isik, U., Phansalkar, N., Giri, R., Helwani, K., and Krishnaswamy, A. (2020). "A Perceptually-Motivated Approach for Low-Complexity, Real-Time Enhancement of Fullband Speech," Proc. Interspeech 2020, 2482-2486.

[4] Schröter, H., Rosenkranz, T., and Maier, A. (2022). "DeepFilterNet: A Low Complexity Speech Enhancement Framework for Full-Band Audio based on Deep Filtering," In 2022 IEEE International Conference on Acoustics, Speech and Signal Processing (ICASSP), paper 9166.

[5] Lv, S., Fu, Y., Xing, M., Sun, J., Xie, L., Huang, J., Wang, Y., and Yu, T. (2022). "S-DCCRN: Super Wide Band DCCRN with learnable complex feature for speech enhancement," In 2022 IEEE International Conference on Acoustics, Speech and Signal Processing (ICASSP), paper 4106.

[6] Le, X., Chen, H., Chen, K., and Lu, J. (2021). "DPCRN: Dual-Path Convolution Recurrent Network for Single Channel Speech Enhancement," Proc. Interspeech 2021, 2811-2815, doi: 10.21437/Interspeech.2021-296

[7] Reddy, C.K.A., Dubey, H., Koishida, K., Nair, A., Gopal, V., Cutler, R., Braun, S., Gamper, H., Aichner, R., and Srinivasan, S. (2021). "INTERSPEECH 2021 Deep Noise Suppression Challenge," Proc. Interspeech 2021, 2796-2800, doi: 10.21437/Interspeech.2021-1609

[8] Vaswani, A., Shazeer, N., Parmar, N., Uszkoreit, J., Jones, L., Gomez, A. N., Kaiser, L., and Polosukhin, I. (2017). "Attention is all you need," Advances in neural information processing systems, 30.

[9] Davis, S. and Mermelstein, P. (1980). "Comparison of parametric representations for monosyllabic word recognition in continuously spoken sentences," IEEE Trans. Acoust., Speech, Signal Process. 28(4), 357-366.

[10] Skowronski, M. D. and Harris, J. G. (2004). "Exploiting independent filter bandwidth of human factor cepstral coefficients in automatic speech recognition," J. Acoust. Soc. Am. 116(3), 1774-1780.

[11] Li, A., Zheng, C., Peng, R., and Li, X. (2021). "On the importance of power compression and phase estimation in monaural speech dereverberation," JASA Express Letters, 1(1), 014802.

[12] Veaux, C., Yamagishi, J., and MacDonald, K. (2017). "CSTR VCTK Corpus: English Multi-speaker Corpus for CSTR Voice Cloning Toolkit," Technical report. The University of Edinburgh. The Centre for Speech Technology Research (CSTR). https://doi.org/10.7488/ds/1994.

[13] Honnet, P.E., Lazaridis, A., Garner, P.N., and Yamagishi, J. (2017). "The SIWIS French Speech Synthesis Database – Design and recording of a high quality French database for speech synthesis," Idiap.

[14] Thiemann, J., Ito, N., and Vincent, E. (2013, June). "The Diverse Environments Multi-channel Acoustic Noise Database (DEMAND): A database of multichannel environmental noise recordings," In Proceedings of Meetings on Acoustics ICA2013 (Vol. 19, No. 1, p. 035081). Acoustical Society of America.

[15] Dean, D., Sridharan, S., Vogt, R., and Mason, M. (2010). "The QUT-NOISE-TIMIT corpus for evaluation of voice activity detection algorithms," In Proceedings of the 11th Annual Conference of the International Speech Communication Association (pp. 3110-3113). International Speech Communication Association.

[16] Ko, T., Peddinti, V., Povey, D., Seltzer, M. L., and Khudanpur, S. (2017, March). "A study on data augmentation of reverberant speech for robust speech recognition," In 2022 IEEE International Conference on Acoustics, Speech and Signal Processing (ICASSP), (pp. 5220-5224). IEEE.

[17] Mysore, G. J. (2014). "Can we automatically transform speech recorded on common consumer devices in real-world environments into professional production quality speech?—a dataset, insights, and challenges," IEEE Signal Processing Letters, 22(8), 1006-1010.

[18] Saki, F., Sehgal, A., Panahi, I., and Kehtarnavaz, N. (2016, March). "Smartphone-based real-time classification of noise signals using subband features and random forest classifier," In 2016 IEEE international conference on acoustics, speech and signal processing (ICASSP) (pp. 2204-2208). IEEE.

[19] Valentini-Botinhao, C., Wang, X., Takaki, S., and Yamagishi, J. (2016, September). "Investigating RNN-based speech enhancement methods for noise-robust Text-to-Speech," in 9th ISCA Speech Synthesis Workshop, pp. 146–152.

[20] Valin, J. M. (2018, August). "A hybrid DSP/deep learning approach to real-time full-band speech enhancement," In 2018 IEEE 20th international workshop on multimedia signal processing (MMSP) (pp. 1-5). IEEE.